\documentclass[%
 aip,
 amsmath,amssymb,
 reprint,%
floatfix]{revtex4-1}

\usepackage{graphicx}
\usepackage{dcolumn}
\usepackage{bm}

\usepackage[utf8]{inputenc}
\usepackage[T1]{fontenc}
\usepackage{mathptmx}
\usepackage{etoolbox}
\usepackage{siunitx}
\DeclareSIUnit\torr{Torr}
\usepackage{booktabs}
\usepackage{comment}
\makeatletter
\def\@email#1#2{%
 \endgroup
 \patchcmd{\titleblock@produce}
  {\frontmatter@RRAPformat}
  {\frontmatter@RRAPformat{\produce@RRAP{*#1\href{mailto:#2}{#2}}}\frontmatter@RRAPformat}
  {}{}
}%
\makeatother
\begin{document}

\preprint{AIP/123-QED}

\title{Autonomous sputter synthesis of thin film nitrides with composition controlled by Bayesian optimization of optical plasma emission}
\affiliation{Materials Science Center, National Renewable Energy Laboratory (NREL), Golden CO, 80401, USA}
\author{Davi M. Febba}
\author{Kevin R. Talley}
\author{Kendal Johnson}
\author{Stephen Schaefer}
\author{Sage R. Bauers}
\author{John S. Mangum}
\author{Rebecca W. Smaha}
\author{Andriy Zakutayev}
\email[Authors to whom correspondence should be addressed: ]{Andriy.Zakutayev@nrel.gov, DaviMarcelo.Febba@nrel.gov}


\begin{abstract}
Autonomous experimentation has emerged as an efficient approach to accelerate the pace of materials discovery. Although instruments for autonomous synthesis have become popular in molecular and polymer science, solution processing of hybrid materials and nanoparticles, examples of autonomous tools for physical vapor deposition are scarce yet important for the semiconductor industry. Here, we report the design and implementation of an autonomous workflow for sputter deposition of thin films with controlled composition, leveraging a highly automated sputtering reactor custom-controlled by Python, optical emission spectroscopy (OES), and a Bayesian optimization algorithm. We modeled film composition, measured by x-ray fluorescence, as a linear function of emission lines monitored during the co-sputtering from elemental Zn and Ti targets in N\textsubscript{2} atmosphere. A Bayesian control algorithm, informed by OES, navigates the space of sputtering power to fabricate films with user-defined composition, by minimizing the absolute error between desired and measured emission signals. We validated our approach by autonomously fabricating Zn\textsubscript{x}Ti\textsubscript{1-x}N\textsubscript{y} films with deviations from the targeted cation composition within relative \SI{3.5}{\percent}, even for \SI{15}{\nano\meter} thin films, demonstrating that the proposed approach can reliably synthesize thin films with specific composition and minimal human interference. Moreover, the proposed method can be extended to more difficult synthesis experiments where plasma intensity depends non-linearly on pressure, or the elemental sticking coefficients strongly depend on the substrate temperature.

\end{abstract}

\maketitle

\section{\label{sec:introduction}Introduction}
Advances in robotics, machine-learning and data science are driving progress in materials science in a data-driven approach, considered as the fourth scientific paradigm \cite{himanen2019} after experimental, theoretical, and computational approaches. With the rise of self-driving laboratories, synthesis and characterization of materials can now be carried out with minimal human intervention and at a faster pace, due to efficient exploration of vast spaces of experimental variables by decision-making algorithms\cite{abolhasani2023,montoya2022,szymanski2021b,stach2021,stein2019}.

Contrary to solution-processed hybrid materials, autonomous synthesis of inorganic thin films by PVD is rather scarce \cite{szymanski2021b}, especially in sputtering, despite its widespread use in research and industry. Some of the few existing reports of autonomous PVD include synthesis and optimization of SrRuO\textsubscript{3}\cite{wakabayashi2019,wakabayashi2022} and TiN\textsubscript{2}\cite{ohkubo2021} thin films prepared by MBE, and Nb-doped TiO\textsubscript{2} \cite{shimizu2020} thin films prepared by sputtering, which is so far the only work reporting a fully autonomous closed-loop PVD synthesis instrument with in-situ measurements and feedback to a control algorithm.

Although these studies focused on the optimization of material properties such as resistivity and crystallinity, precise control of cation and anion composition in inorganic thin films is of paramount importance. For example, it has been theoretically predicted \cite{pan2020} and experimentally \cite{melamed2022} demonstrated that short-range ordering tunes the optical absorption edge in the long-range disordered alloy (ZnSnN)\textsubscript{1-x}(ZnO)\textsubscript{2x} at a very specific composition of $x=0.25$. Also, the resistivity and bandgap of ternary nitrides and their alloys depend mostly on cation composition \cite{greenaway2022,febba2021}. Furthermore, many promising oxynitrides were recently predicted to possess semiconductor electrical transport and ferroelectric polarization properties similar to halide perovskites, but with longer-term stability \cite{hartman2020}. However, the specific O:N = 2:1 required for these properties is challenging to control during synthesis \cite{heinselman2022}.

In this context, here we report the design and implementation of an autonomous instrument for controlling the composition of thin films with minimal human intervention, leveraging a highly automated sputtering reactor and optical emission spectroscopy (OES). By fabricating Zn\textsubscript{x}Ti\textsubscript{1-x}N\textsubscript{y} thin films with simultaneous monitoring of optical emission lines from the sputtering of elemental targets, we show that cation composition, spanning a wide range, can be expressed as a function of emission lines only. 

Informed by OES measurements, a closed-loop control algorithm with Bayesian optimization as its decision-making agent can effectively optimize the power on each sputtering source to fabricate thin films with specific cation composition, defined prior to deposition, with minimal human interference. Moreover, we show that our model can accurately predict film composition regardless of total power or gas flow, as long as the OES signal is reproduced, but re-calibration is needed if autonomous depositions are carried out at chamber pressures different than that set during model calibration. 

\section{\label{sec:sputtering}Overview of the sputtering instrument}

To enable autonomous synthesis of thin films with controlled composition, we recently designed and built a highly-automated sputtering reactor. Equipped with four sputtering sources (cathodes), and able to source RF and DC power to the substrate, this high-vacuum ($< \SI{1e-7}{Torr}$) instrument allows the exploration of a wide substrate temperature range, from cryogenic temperatures up to 1000 °C.

\begin{figure}
    \centering
    \includegraphics[width=\columnwidth]{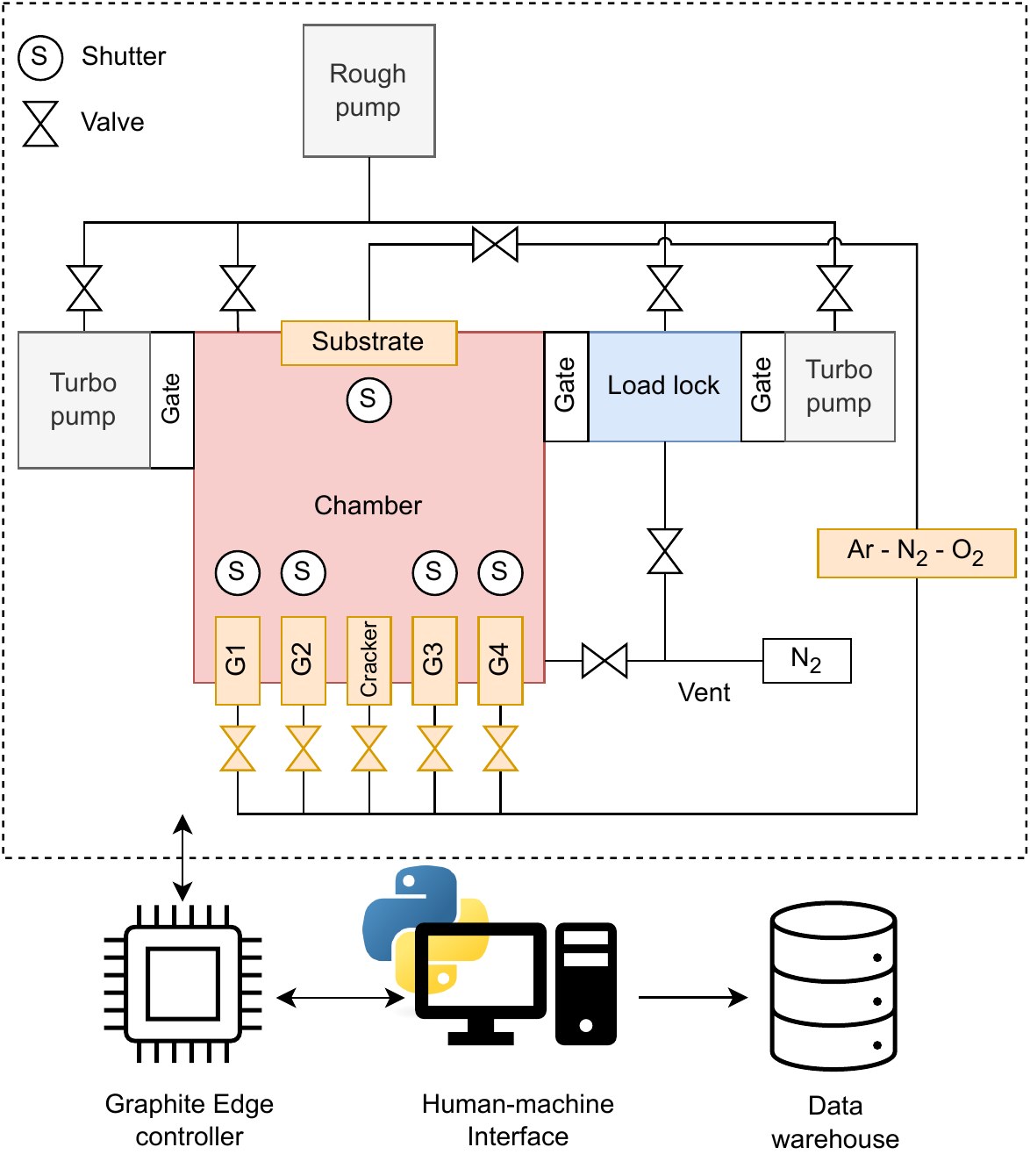}
    \caption{Diagram of the autonomous sputtering system, showing gas distribution, sputtering sources and data flow. A Python script controls all chamber parameters by interacting with a Graphite Edge controller, and real-time deposition data are recorded and stored in a data warehouse.}
    \label{fig:schematics}
\end{figure}

Fig. \ref{fig:schematics} shows a high-level diagram of this system, depicting sputtering sources, distribution of process gas and data flow. Additional capabilities include control of gas mixing and distribution to key locations within the sputtering environment, such as individual targets and substrate, time-sequenced shutters, and pressure control via positioning of the gate valve to the turbo molecular pump.

The high degree of automation of this sputtering instrument was accomplished by connecting all its sensors and actuators, such as pressure gauges, power supplies, and pneumatic valves, to a Graphite Edge controller (Red Lion Controls) that interfaces with the user through a Human-Machine Interface (HMI), developed with Crimson 3.1. 

This approach allows the implementation of customized software solutions and enables the user to execute complex programmable synthesis recipes, besides offering room for further custom automation. Moreover, real-time deposition data of sputtering parameters (such as power, voltage, pressure, gas flow) are recorded and loaded into a data warehouse \cite{talley2021}.

A unique feature of this reactor is that the Graphite Edge controller supports the Open Platform Communications Unified Architecture protocol (OPC UA), acting as a server. In this way, a Python client can directly communicate with the controller through Python OPC UA libraries, effectively controlling all chamber parameters.

Thus, Python libraries for machine learning and optimization can be leveraged to control sputtering parameters in real-time, which is usually not possible for chambers and accompanying systems available, making this instrument an excellent platform for implementing an autonomous workflow for the synthesis of thin films. 

\section{\label{sec:oes}Optical Emission Spectroscopy}

Optical Emission Spectroscopy (OES) is a passive optical diagnostic method that analyzes light emitted from excited atoms and molecules in a plasma environment and, due its simplicity and non-intrusive aspect, it was one of the earliest techniques applied to the analysis of sputtering plasma \cite{britun2020}.

The basic physical process in OES is the excitation of particles by electron impact from level \textit{i} to \textit{j} and decay into level \textit{k} by spontaneous emission with transition probability \textit{A\textsubscript{jk}}, resulting in a line emission \textit{$\epsilon$\textsubscript{jk}} with wavelength $\lambda = hc/(E_{j}-E_{k})$, where h is the Planck's constant, c is the speed of light, and E\textsubscript{j} and E\textsubscript{k} are the energy levels of states \textit{j} and \textit{k}, respectively. 

This wavelength is detected by the emission spectrometer and is a fingerprint for the radiating particle \cite{fantz2006}, while line intensities are proportional to the density of particles in the plasma \cite{clenet1997}. It has been shown by previous works that the intensity of emission lines monitoring during sputtering can be connected to resulting film composition \cite{shin2018,wu2002} and material properties \cite{rachdi2021,salimian2022,yang2021}, besides being useful for process control \cite{drury2021,posada2015}.

In glow discharge optical emission spectroscopy (GDOES), a sample of interest acts as a cathode and is sputtered, usually with Ar. By analyzing the emitted light with a spectrometer, elemental composition of the sample and depth profile analysis can be carried out, since the basic assumption is that the intensity of a certain emission line of an element is a function of the concentration of this element in the analyzed material \cite{weiss2006}.

Since it is a comparative technique, it needs calibration \cite{weiss2015}: reference materials of known composition must be analyzed in order to build a calibration function $I_{\lambda(X),M} = f(c_{E,M})$, where $X$ is the element to be detected in the material $M$ and $I_{\lambda(X)}$ is the line intensity at wavelength $\lambda$ of the element $X$ with concentration $c$ in the material $M$. 

\section{\label{sec:calibration}Calibration}

Optical emission spectroscopy can measure film composition after a calibration procedure is performed, which consists in finding a relation between intensity of emission lines of interest and film composition, which must be measured by a direct technique. In this section, we will discuss each part of this procedure: fabrication of thin films and composition measurement by Energy Dispersive X-ray Fluorescence (EDXRF), OES characterization, and establishing a connection between EDXRF and OES measurements through an analytical function.

\subsection{\label{sec:fabrication}Thin film fabrication}

To demonstrate autonomous synthesis of thin films with controlled composition, Zn\textsubscript{x}Ti\textsubscript{1-x}N\textsubscript{y} was considered as a validation case as it has been successfully synthesized by our group \cite{greenaway2022} in the sputtering instrument described in section \ref{sec:sputtering}. In this work, we assume nitrogen composition will fall on the TiN - ZnTiN\textsubscript{2} - Zn\textsubscript{3}N\textsubscript{2} tie lines, although it is not explicitly measured since cation composition is our focus.

We fabricated thin films on glass substrates (2" x 2" Corning Eagle XG Glass). Before each deposition, the chamber was evacuated at pressures lower than \SI{1e-7}{Torr}, after transferring the substrate from a load lock chamber. Two-inch diameter Zn (\SI{99.99}{\percent} purity) and Ti (\SI{99.995}{\percent} purity) elemental targets were excited by radio-frequency (RF) sputtering sources. Common to all depositions, the chamber pressure was set at \SI{10}{\milli Torr} after introducing 20 sccm of Ar adjacent to each sputtering source (40 sccm total) and 20 sccm of N\textsubscript{2} adjacent to the substrate gas inlet. 

\begin{figure}
    \centering
    \includegraphics[width=\columnwidth]{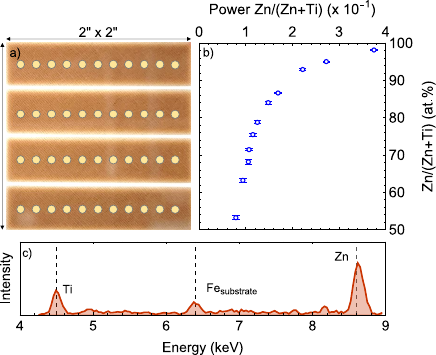}
    \caption{(a) Picture of a sample fabricated on a 2" x 2" glass substrate, with approximate positions of EDXRF measurements. Diagonal striations are from a cotton cloth underneath the substrate. (b) Average cation composition as a function of power ratio on Zn and Ti targets, obtained from EDXRF measurements across 44 points on the film. (c) Example of an EDXRF spectrum measured at a single spot, showing Zn and Ti  K$\alpha$ spectral lines used for composition analysis, and an Fe line from the substrate.}
    \label{fig:sample}
\end{figure}

Depositions lasted between \SI{45}{\min} and \SI{2}{\hour}, always after a pre-sputtering step of \SI{30}{\min} with the substrate shutter closed to clean the surface of the targets, and no substrate heating or cooling was applied. The substrate was rotated during deposition to grow films with homogeneous composition. Fig. \ref{fig:sample}a shows a sample of zinc titanium nitride fabricated by this process.

Several depositions were carried out with varying powers applied to each target. The resulting films spanned a wide composition, expressed as the average of Zn/(Zn+Ti) (at.\%) in Fig \ref{fig:sample}b, obtained by EDXRF measurements taken at 44 distinct spots on the film (Fig. \ref{fig:sample}a) with a Fischerscope Energy Dispersive X-ray Fluorescence (EDXRF) and accompanying WinFTM analysis software. Average thickness, estimated from EDXRF (calibrated with ellipsometry measurements\cite{greenaway2022}) was between \SI{18.0(2)}{\nano\meter} and \SI{110.0(3)}{\nano\meter} for all depositions. Uncertainty in composition and thickness was taken as two times the standard uncertainty, i.e., $2s/\sqrt{n}$, where $s$ is the standard deviation of 44 measurements ($n$), resulting in a confidence interval (C.I.) of approximately \SI{95}{\percent}.

The range of power applied to each sputtering source takes into account the maximum power allowed on each target, as well as the minimum to sustain a plasma. Thus, powers ranging from \SIrange{12}{30}{\watt} and \SIrange{50}{140}{\watt} for the Zn and Ti targets, respectively, resulted in samples with average composition spanning the range \SIrange{53}{98}{\percent}, as summarized in Fig. \ref{fig:sample}. Although it could be possible to obtain cation compositions below \SI{50}{\percent} Zn/(Zn+Ti) (at.\%), the power on the Zn target would be too small, below \SI{12}{\watt}, which was seem to result in films with irreproducible composition for the same chamber conditions. On the other hand, the power on the Ti target would be at the maximum of \SI{150}{\watt}, and could not be increased without risking damage to the target.

\subsection{\label{sec:oes characterization}Plasma monitoring}

\begin{figure}
    \centering
    \includegraphics[width=\columnwidth]{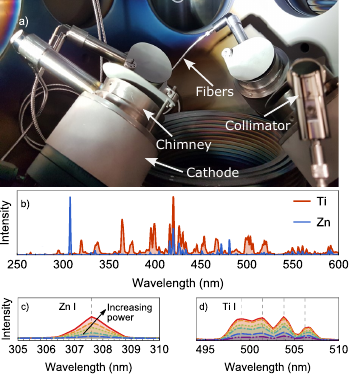}
    \caption{a) Internal view of the process chamber, showing 2-inch chimneys adapted to accommodate \SI{90}{\degree} collimators to collect emission spectra from each sputtering source. b) Optical emission spectra from Zn and Ti targets sputtered with Ar, between \SI{250}{\nano\meter} and \SI{600}{\nano\meter}, where most of the Ti I and Zn I lines are found. Increase of Zn I (c) and Ti I (d) emission lines for increasing sputtering power on each target.}
    \label{fig:combi9}
\end{figure}

To enable in-vacuum plasma monitoring with OES in our sputtering system, 2-inch stainless steel chimneys were specially adapted to accommodate \SI{90}{\degree} collimators with \SI{3}{\degree} acceptance angle, as shown in Fig. \ref{fig:combi9}a, each monitoring the glow discharge just above the target.

During each deposition, optical emission spectra from each target were independently recorded by a multi-channel Plasus Emicon MC spectrometer, between \SI{195}{\nano\meter} and \SI{1105}{\nano\meter}, avoiding signal interference during co-sputtering. A useful feature of this setup is that each collimator is equipped with a coating-protection device, consisting of a quartz plate and a capillary cartridge, to  avoid deposition of sputtered material on the collimator optics. 

Each spectrometer integration time was set to \SI{50}{\milli\second}, and 10 spectra were collected and averaged to increase the signal-to-noise ratio, amounting to a total recording time of about \SI{500}{\milli\second} per spectrum. Background OES spectra were collected before igniting the plasma on the sputtering sources and then subtracted from all subsequent measurements.

Spectrum measurement and signal processing were carried out with the Plasus Emicon MC software that accompanies the spectrometer, while analysis of emission lines was accomplished with the Specline software, which has an extensive database of spectral lines for atoms, ions and molecules, and features automated line detection based on the elements of interest.

To detect non-overlapping Ti and Zn lines that are sensitive to changes in sputtering power, the Ti and Zn targets were initially sputtered with Ar at \SI{10}{\milli Torr}, by injecting 20 sccm of Ar through a gas inlet adjacent to each sputtering source. Fig. \ref{fig:combi9}b shows that a Zn I line at \SI{307.5}{\nano\meter} and Ti I lines between \SI{496}{\nano\meter} and \SI{510}{\nano\meter} are suitable choices, since they do not show any overlap and are intense enough to be detected with the aforementioned spectrometer acquisition settings.

Moreover, these lines are also sensitive to changes in power, as shown in Figs. \ref{fig:combi9}c-d. However, to avoid inaccurate signal evaluation due to broadening and peak shift, common in OES analysis \cite{ley2014}, the integral under the emission signal was taken instead of line intensities. For simplicity, we will refer to this integral as intensity only.

Thus, to incorporate information from both sputtering targets into the analysis, the normalized ratio was taken, as given by Eq. (\ref{ratio}):

\begin{equation}
    I_{OES} = \frac{I_{Zn}}{I_{Zn}+I_{Ti}},\ I_{Zn} = \int_{306}^{310} I \,d\lambda,\ I_{Ti} = \int_{496}^{510} I \,d\lambda,
    \label{ratio}
\end{equation}
where $I$ is the intensity of the optical emission signal monitored by independent channels over each target. Then, considering that $I_{OES}$ was observed to be stable over time, the median of $I_{OES}$ was taken between signal stabilization -- \SI{10}{\min} from plasma ignition -- and the end of deposition, so that a single OES parameter characterizes the plasma environment.

\subsection{\label{sec:analytical}Analytical function}

In GDOES, the goal is initially to find a calibration function: a model that describes emission lines as a function of elemental composition in reference samples of known composition \cite{weiss2015}. Then, to characterize samples of unknown composition, this calibration function is mathematically inverted, and the resulting model is called the analytical function. By measuring the intensity of a specific line of element $X$, its composition in a material $M$ can be obtained \cite{weiss2023}.

\begin{figure}
    \centering
    \includegraphics[width=\columnwidth]{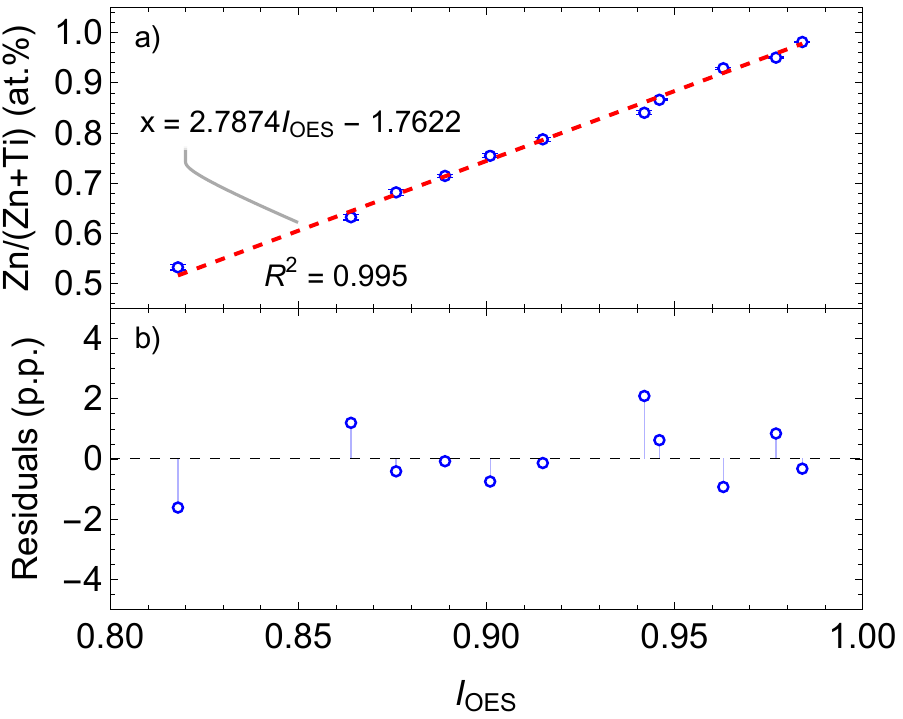}
    \caption{Linear fit of Zn/(Zn+Ti) (at.\%) in Zn\textsubscript{x}Ti\textsubscript{1-x}N\textsubscript{y} as a function of OES signal (a) and corresponding residuals (b), expressed as percentage points (p.p.).}
    \label{fig:model}
\end{figure}

Therefore, in our approach, we establish an analytical function through the calibration procedure, i.e., fabricating samples and measuring their composition with EDXRF, and then finding a model to express film composition as a function of the normalized ratio of intensity of emission lines given by Eq. (\ref{ratio}). Fig. \ref{fig:model} shows that a simple linear relation between $I_{OES}$ and composition was obtained, with $R^2=0.995$, Root Mean Square error of \num{1.0e-2}, and maximum residual of 2.1 percentage points.

All depositions and OES measurements taken to find this analytical function were carried out on different days, spanning several weeks, including many sources of random changes, such as venting the chamber to perform routine maintenance work, inspection of the collimators for possible coating, and so on. Thus, the analytical function incorporates all these changes that occurred over time, making the model more robust against these sources of noise. However, for future experiments, all the steps for the calibration procedure can be completed in about one day to avoid these sources of noise.

\section{\label{sec:autonomous}Control algorithm}

\begin{figure}
    \centering
    \includegraphics[width=\columnwidth]{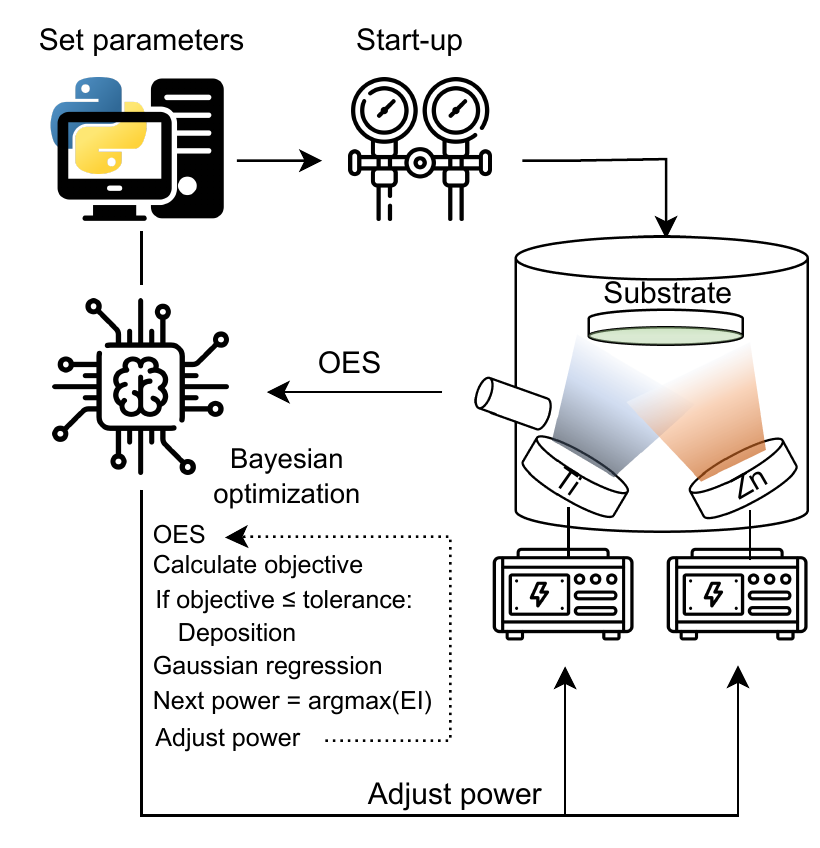}
    \caption{Flowchart of the autonomous sputtering workflow. After automatic chamber start-up, a Bayesian optimization algorithm controls the sputtering powers based on the analysis of optical emission signals from each target. The only step requiring human intervention is to set up initial chamber and optimization parameters.}
    \label{fig:flowchart}
\end{figure}

An autonomous closed-loop control system leveraging the high degree of automation of our sputtering reactor was implemented using Python. The flowchart in Fig. \ref{fig:flowchart} summarizes the workflow. At its core, a Bayesian optimization algorithm, implemented with the scikit-optimize (skopt) \cite{head2022} package, controls the radio frequency sputtering power on elemental targets (Zn and Ti, in this study) and, informed by OES measurements, explores the 2D-space of power applied on the targets until a user-defined OES signal $(I_{OES})$ that corresponds to a specific film composition is obtained.

The goal of the optimization algorithm is to minimize an objective function, which in this case is the absolute error between the measured $I_{OES}$ and the user-defined $I_{OES}$ (setpoint), given by Eq. (\ref{eq:error}):

\begin{equation}
objective = \left|I_{OES}^{current} - I_{OES}^{setpoint}\right|,
\label{eq:error}
\end{equation}
where $I_{OES}^{current} = median(I_{OES})$ over the last \SI{10}{\second} of data, to filter out measurement noise.

A Gaussian process was employed as surrogate model for the unknown function that models Eq. (\ref{eq:error}) as a function of the sputtering powers on the targets. The Radial Basis Function (squared-exponential) was used as kernel, with its hyperparameters optimized at each optimization loop by scikit.

After loading the substrate into the chamber, the only step requiring human intervention is to define start-up optimization and chamber parameters, such as process pressure, gas flows and power on each target, and to specify the desired OES signal (setpoint). Upon setting these initial parameters, the control algorithm initially checks for high vacuum condition in the sputtering chamber. If this condition is satisfied, an automatic routine ignites the plasma on each sputtering source, adds reactive gas (N\textsubscript{2}) through the substrate gas inlet and controls the chamber pressure.

A countdown for OES signal stabilization begins, and, after it completes, the Bayesian optimization loop starts. Starting with an arbitrary pair of sputtering powers on the two targets set during start-up, the optimization algorithm checks if $I_{OES}$ is stable by verifying it is within a maximum allowed deviation of \SI{3}{\percent} from its median over the last \SI{10}{\second} of data. Then, if the signal is stable, it measures $I_{OES}^{current}$ and calculates the objective function given by Eq. (\ref{eq:error}).

If the objective is higher than a tolerance criterion, a Gaussian process is then fitted to the data, and the point where the Expected Improvement (EI) acquisition function is maximized is then suggested as the next sampling point, within a search space: a new pair of sputtering powers are then set on the RF power supplies connected to the sputtering sources, and adjusted by a ramp procedure at a rate of \SI{2}{\watt\per\second} (user-defined) to avoid plasma destabilization due to fast changes in power.

The algorithm then waits for signal stabilization and takes another $I_{OES}^{current}$ measurement. This whole process -- OES analysis, evaluation of the objective function, Gaussian regression, sampling a next point -- continues until convergence is achieved, i.e., the objective function falls below a tolerance criterion, or a maximum number of iterations is reached, as depicted in Fig. \ref{fig:flowchart}. If the convergence criterion is satisfied, the substrate shutter is automatically opened, and deposition starts.

Although the sputtering powers could be controlled by a more conventional Proportional-Integral-Derivative (PID) approach\cite{wakabayashi2023} to minimize the objective function, human tuning of PID parameters to control the growth variables represent a serious bottleneck for an autonomous framework, especially if more variables must be controlled, hindering scalability and maintainability, since re-tuning of these parameters is often needed. Conversely, a control method based on Bayesian optimization eliminates these issues by efficiently sampling the search space and refining the surrogate model in a data-driven approach: our knowledge about the likely objective function improves as more data are observed \cite{shahriari2016}.

Note that this optimization problem is a degenerate one: several sets of power levels can lead to the same OES signal. As we will discuss later in section \ref{sec:degeneracy}, our approach of taking the first set of powers for which the algorithm finds convergence does not lead to any detriment in the prediction results, since as long as the desired OES signal is obtained, the composition will not significantly change.

\begin{figure*}
    \centering
    \includegraphics[width=\textwidth]{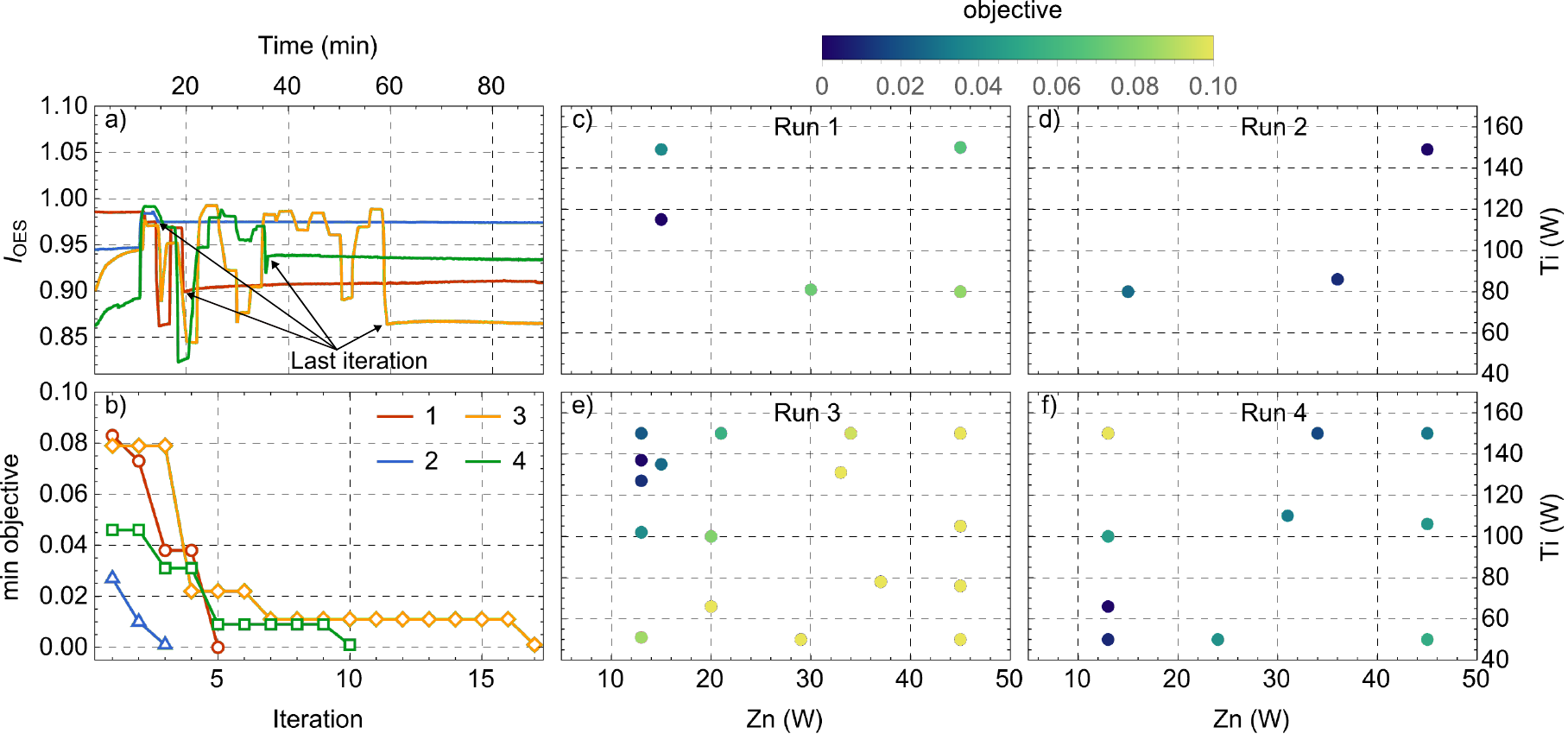}
    \caption{Optimization results for 4 validation runs. OES signal as a function of time (a), while the Bayesian optimization algorithm searches for the optimal sputtering powers. The convergence plot (b) shows the evolution of the minimum of the objective function, while (c)-(f) show the sputtering powers visited by the algorithm for each validation run, with colors representing the value of the objective function.}
    \label{fig:convergence}
\end{figure*}

\section{\label{sec:oresults}Results}
\subsection{\label{sec:validation}Validation}

To validate the proposed approach, the aim was to fabricate films with a specific composition, defined prior to deposition: \SI{75}{\percent}, \SI{95}{\percent}, \SI{65}{\percent} and \SI{85}{\percent} Zn/(Zn+Ti) (at.\%), denoted as validation runs 1, 2, 3, and 4, respectively, in Table \ref{tab:experiments}. Zn-rich conditions were chosen due to irreproducible OES measurements and film composition obtained at low powers on the Zn target, as previously mentioned, and the search range for the Bayesian algorithm was set as \SIrange{50}{150}{\watt} for the Ti target and \SIrange{13}{45}{\watt} for the Zn target.

For these experiments, the sputtering chamber was kept at \SI{10}{\milli Torr} by adding 20 sccm of Ar through each sputtering source and 20 sccm of N\textsubscript{2} through the substrate, as in the calibration procedure. We then set \SI{10}{\min} as initial stabilization window after automatic plasma ignition, which was reduced to \SI{2}{\min} after ramping the power on the sputtering sources during the optimization loop.

After chamber start-up with arbitrary sputtering powers, the control system adjusts the powers to obtain OES signals that would result in films with desired composition. The algorithm was allowed to run for up to 30 iterations, and the convergence threshold was set as \num{6e-3}, which was found to guarantee convergence and resulted in a good trade-off between convergence time and accuracy of the predicted film composition.

\begin{table}
\centering
\caption{Results of the validation experiments, with relative errors between predicted and actual film composition, expressed as Zn/(Zn+Ti) (at.\%) and averaged across 44 distinct spots on the substrate.}
\label{tab:experiments}
\begin{ruledtabular}
\begin{tabular}{ccccc}
Run & Goal (\%) & Predicted (\%) & Actual (\%) & Prediction error (\%) \\ \hline
1 & 75 & 75.0 & \num{76.9(04)} & -2.5 \\
2 & 95 & 95.3 & \num{92.2(01)} & 3.4 \\
3 & 65 & 64.6 & \num{65.5(03)} &  -1.4 \\
4 & 85 & 85.3 & \num{86.3(06)} &  -1.2 \\ 
\end{tabular}
\end{ruledtabular}
\end{table}

Fig. \ref{fig:convergence}a shows the evolution of the emission signal as a function of time, while the Bayesian optimization algorithm searches for the best set of powers to achieve the desired setpoint signal. After convergence is achieved, the OES signal is stable over time until the deposition ends. 

Although convergence was fast for Runs 1 and 2, as shown in Fig. \ref{fig:convergence}b, with only 5 and 3 iterations needed for convergence, respectively, the algorithm sometimes needs more iterations due its stochastic nature, as seen for Run 3, which took 17 iterations. Nonetheless, the algorithm was able to achieve convergence for all four validation runs. Moreover, Fig. \ref{fig:convergence}c-f show the values of sputtering powers explored by the Bayesian optimization algorithm and the resulting objective function for all the validation runs.

\begin{figure*}
    \centering
    \includegraphics[width=\textwidth]{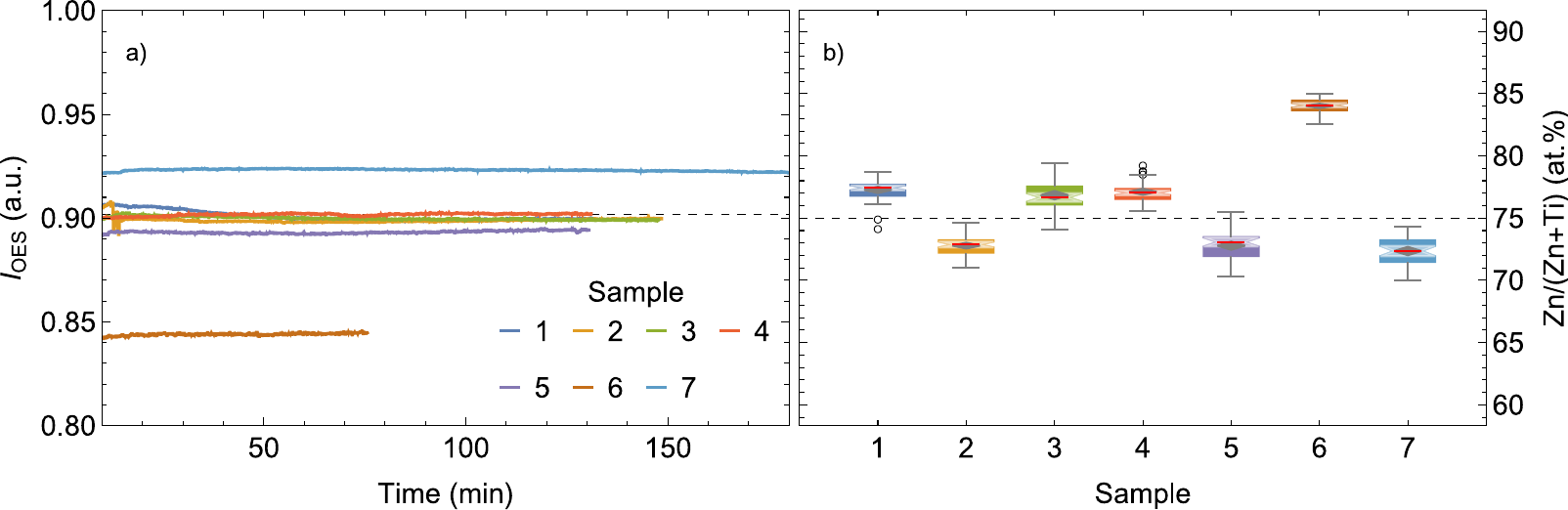}
    \caption{Effects of total power (samples 1, 2, and 3), gas flow (samples 1, 4, and 5) and pressure (samples 6 and 7), on the OES signal (a) and film composition measured by EDXRF (b). Diamonds edges and notches denote the 95 \% C.I. for the mean (center of gray diamonds) and median (red dash) in the notched box plots, respectively.}
    \label{fig:effects}
\end{figure*}

To verify whether the resulting films had the targeted compositions, EDXRF measurements were carried out across 44 spots on the substrate, and the average value was taken as film composition, as previously discussed. Since the algorithm had a convergence threshold for the OES signal taken as the optimum (convergence happens for when $objective\leq \num{6e-3}$), we report the predicted composition, i.e., the composition predicted from the analytical function from Fig. \ref{fig:model} based on the OES measurement taken at the last iteration of the optimization routine. Average film thickness was \SI{26.7(2)}{\nano\meter}, \SI{141.6(4)}{\nano\meter}, \SI{41.6(3)}{\nano\meter} and \SI{15.4(2)}{\nano\meter}, for samples 1, 2, 3, and 4, respectively.

Prediction errors, calculated as $(predicted/actual-1)$, express the relative error between predicted and actual film composition. As shown in Table \ref{tab:experiments}, the proposed approach was able to autonomously fabricate thin films with composition within \SI{3.5}{\percent} of the desired value. As more depositions are carried out with simultaneous plasma monitoring, the resulting OES signal and film composition can be retrofitted into the analytical function to improve its accuracy in predicting film composition. Other routes of improvement include the fabrication of thicker films for higher precision of EDXRF measurements, waiting a longer time for OES signal stabilization after a new power set is suggested by the algorithm, and reducing the convergence criterion from \num{6e-3}, at the expense of convergence time.

\subsection{\label{sec:degeneracy}Degeneracy: total power effects}

We studied if degeneracy would have any adverse effects in the prediction results. As previously mentioned, the optimization problem of finding the optimal set of sputtering powers to achieve a desired OES signal is a degenerate one: several sets of powers may result in the same signal.

For this study, we varied the power applied to each target for three independent depositions while keeping $I_{OES}\approx0.902$ (dashed line on Fig. \ref{fig:effects}a), which corresponds to a 75 \% Zn/(Zn+Ti) (at.\%) (dashed line on Fig. \ref{fig:effects}b) according to the analytical function. The chamber pressure was set at \SI{10}{\milli Torr} by adding 20 sccm of Ar through gas inlets adjacent to each sputtering source (total of 40 sccm), and 20 sccm of N\textsubscript{2} adjacent to the substrate, as summarized in Table \ref{tab:effects} (samples 1, 2, and 3). Fig. \ref{fig:effects}a shows that it was possible to reproduce the desired OES signal for three different sets of power levels, with a maximum deviation of \num{3.3e-3} between the desired $I_{OES}$ and the median of $I_{OES}$ taken from \SI{10}{\min} to the end of the deposition. As seen in Fig. \ref{fig:effects}b, film composition did not show any significant changes, even though the powers on both targets changed. Prediction errors between actual and predicted composition (given by the analytical function from Fig. \ref{fig:model} based on the median of the OES signal between \SI{10}{\min} and the end of each deposition), were \SI{-3.8}{\percent}, \SI{1.7}{\percent} and \SI{-3.3}{\percent}, respectively.

Therefore, this degeneracy caused by simultaneously optimizing two power supplies does not constitute an issue as long as the OES signal can be reproduced, since the prediction errors are of the same order of those found for the validation experiments shown in Table \ref{tab:experiments}. However, other film properties may change, such as morphology and crystallinity, but since we are only optimizing for composition, these side effects are out of scope here.

\subsection{\label{sec:chamber effects}Effects of chamber pressure and gas flow}

\begin{table}
\centering
\caption{Different total powers and gas flows of Ar and N\textsubscript{2} do not cause large relative errors when predicting film composition based on emission signals, contrary to changes in the chamber pressure, which invalidate the analytical function obtained during calibration.}
\label{tab:effects}
\begin{ruledtabular}
\resizebox{\columnwidth}{!}{%
\begin{tabular}{ccccccc}
Sample & Zn (W) & Ti (W) & Ar (sccm) & N\textsubscript{2} (sccm) & mTorr & Error (\%) \\ \hline
1 & 15 & 115 & 40 & 20 & 10 & -3.8 \\
2 & 17 & 145 & 40 & 20 & 10 & 1.7 \\
3 & 13 & 101 & 40 & 20 & 10 & -3.3 \\
4 & 15 & 115 & 30 & 10 & 10 & -2.9 \\
5 & 15 & 115 & 20 & 15 & 10 & -0.4 \\
6 & 15 & 115 & 40 & 20 & 5 & -30.0 \\
7 & 15 & 115 & 40 & 20 & 15 & 11.7
\end{tabular}}
\end{ruledtabular}
\end{table}

We also investigated the effects of chamber pressure and gas flow on the OES signal and on the accuracy of the analytical function (Fig. \ref{fig:model}) for predicting film composition. For this purpose, we fabricated films at different total flows of Ar and N\textsubscript{2}, and at lower (\SI{5}{\milli Torr}) and higher pressures (\SI{15}{\milli Torr}), as summarized in Table \ref{tab:effects}, while keeping constant the set of sputtering powers that the algorithm found in the validation experiments for a film with 75 \% Zn/(Zn+Ti) (at.\%): \SI{115}{\watt} and \SI{15}{\watt} on the Ti and Zn targets, respectively.

As shown in Fig. \ref{fig:effects}a -- samples 1, 4 and 5 -- the optical emission signal $I_{OES}$ did not show significant changes for different gas flows, with a constant pressure set at \SI{10}{\milli Torr}. For these samples, predicting film composition by feeding $I_{OES}$ into the analytical function (Fig. \ref{fig:model}) results in small prediction errors, expressed as the relative error between actual and predicted composition, as demonstrated in Table \ref{tab:effects}.

Therefore, the proposed approach can still produce reliable results for different gas flows, but only if the chamber pressure is equal to that set during the calibration step: decreasing the pressure to \SI{5}{\milli Torr} resulted in a significant drop in the emission signal with an uncorrelated increase of Zn in the film (Fig. \ref{fig:effects}b - sample 6).

If the analytical function was still valid for this case, \SI{58.8}{\percent} Zn/(Zn+Ti) would be observed in the film, according to the observed emission signal. Fig. \ref{fig:effects}b shows that the composition of Zn/(Zn+Ti) in the resulting film was \SI{84}{\percent}, thus invalidating the analytical function for depositions at pressures lower than that of the calibration procedure.

If this was an autonomous deposition with a desired \SI{75}{\percent} of Zn/(Zn+Ti) in the film, the algorithm would try to converge to the setpoint signal of 0.902 by increasing the power on Zn or decreasing the power on Ti, or both, leading to an even higher percentage of Zn in the film, although the desired OES signal would be achieved.

Higher pressures lead to the same effect: increasing the pressure to \SI{15}{\milli Torr} resulted in a shift of the emission signal towards a higher value (Fig. \ref{fig:effects} - sample 7), with an uncorrelated decrease in Zn/(Zn+Ti) (at.\%) in the film to \SI{72.4}{\percent}. Although a higher pressure did not cause significant changes in actual composition, the predicted composition from the analytical function in Fig. \ref{fig:model} was \SI{80.9}{\percent}, resulting in a relative error close to \SI{12}{\percent}, too large for the purpose of controlling film composition. Again, if this was an autonomous deposition, the algorithm would try to reduce the emission signal, which would lead to even higher prediction errors.

\section{Discussion}

The effects of chamber pressure in the resulting film composition and OES signal (Fig. \ref{fig:effects}) can be explained by two effects\cite{clenet1997}: at low pressures, the mean free path of the sputtered atoms increases, since they do not experience a significant amount of collisions with Ar atoms and do not get accumulated in the discharge, reducing the intensity of the emission signal. 

On the other hand, the substrate is also sputtered by energetic particles, and lighter atoms (Ti, in this case) are preferentially resputtered from the film, which explains the higher Zn/(Zn+Ti) ratio in the film fabricated at low pressure, even though gas flows and powers were the same as for sample 1. In contrast, this phenomenon is reduced at higher pressures, explaining the decreased Zn/(Zn+Ti) ratio at \SI{15}{\milli Torr}.

Any effects at the substrate are thus not taken into account, and substrate temperature can also cause large errors on the prediction results, since substrate heating can result in different film compositions even though the plasma conditions are the same. In this way, the variation caused by substrate effects constitutes a hindrance, since several film compositions could be obtained for the same OES signal, contrary to the previously discussed degeneracy caused by total power.

Nonetheless, the proposed approach is useful for autonomous synthesis of thin films with user-defined composition. Taking advantage of a short calibration procedure and an automated deposition setup, several analytical functions can be quickly obtained for different pressures and substrate temperatures, making it possible to explore chamber conditions for which film quality is enhanced.

Finally, to further improve the acceleration provided by the synthesis of thin films with controlled composition, future work will focus on developing calibration procedures that are independent of chamber conditions and geometry.

\section{Conclusion}

We reported the design and implementation of an autonomous instrument for sputter synthesis of thin films with controlled cation composition, in a highly automated reactor that interfaces with Python scripts. After a calibration procedure to correlate actual film composition, measured by ex-situ Energy Dispersive X-ray Fluorescence (EDXRF), and in-situ optical emission spectroscopy (OES) data obtained during the RF sputtering of elemental targets, we showed that a linear function can predict composition based solely on emission lines. 

Informed by real-time OES measurements, a Bayesian optimization algorithm optimized the RF power applied to sputtering sources to synthesize films with user-defined composition. As a case study, our instrument fabricated Zn\textsubscript{x}Ti\textsubscript{1-x}N\textsubscript{y} thin films targeting x = 0.65, 0.75, 0.85, and 0.95. EDXRF measurements showed that the proposed approach resulted in films with x = 0.65, 0.77, 0.86, and 0.92, thus in good agreement with the targeted composition.

However, for accurate results, it is crucial to maintain the chamber pressure at the level set during calibration, as any variations in this parameter can shift the optical emission signal with uncorrelated changes in the final film composition, leading to substantial inaccuracies. Conversely, changes in gas flow and total sputtering power do not seem to increase the prediction errors if the chamber pressure can be kept at a constant level.

\begin{acknowledgments}
This work was authored by the National Renewable Energy Laboratory, operated by Alliance for Sustainable Energy, LLC, for the U.S. Department of Energy (DOE) under Contract No. DE-AC36-08GO28308. Funding provided by the National Renewable Energy Laboratory (NREL), under the Laboratory Directed Research and Development (LDRD) program. R.W.S. acknowledges support from the Director’s Fellowship within the NREL LDRD program. The views expressed in the article do not necessarily represent the views of the DOE or the U.S. Government. Schematics were designed using icons made by FreePik and Vectoricons from www.flaticon.com.
\end{acknowledgments}

\section*{Author Declarations}

The authors have no conflicts to disclose.

\section*{Data Availability Statement}

The data that support the findings of this study are available from the corresponding author upon reasonable request.

\bibliography{references}

\end{document}